\documentclass[screen,sigplan,10pt,nonacm]
{acmart}

\startPage{1}

\setcopyright{none}

\usepackage{booktabs}   
\usepackage{subcaption} 
\usepackage[figurename=Fig.]{caption}
\usepackage[tablename=Tab.]{caption}
\usepackage{amsmath}
 \usepackage{amssymb}
\usepackage{amsthm}
\usepackage{textcomp}
\usepackage{mathtools}
\usepackage{amscd}
\usepackage{lipsum}
\usepackage{amsfonts}
\usepackage{varwidth}
\usepackage{graphicx}
\usepackage{listings}
\usepackage{float}
\usepackage{mathpartir}
\usepackage{multirow}
\usepackage[noend]{algorithmic}
\usepackage{mathrsfs}
\usepackage{fancyvrb}
\usepackage{float}
\usepackage{wrapfig}
\usepackage{xspace}
\usepackage{enumitem}
\usepackage{natbib}
\usepackage{tkz-base,tikz}
\usepackage{pgfplots}
\usepackage{subcaption}

\makeatletter 
\def\arcr{\@arraycr}
\makeatother

\setlist[itemize]{leftmargin=*}
\setlist[enumerate]{leftmargin=*}

\floatstyle{plaintop}
\restylefloat{table}

\defcitealias{Coq-manual}{Coq}
\defcitealias{liquidity}{Liquidity}
\defcitealias{dao-project}{DAO}
\defcitealias{solidity}{Solidity}
\defcitealias{solidity-bugs}{Known Bugs in Solidity Compiler}
\defcitealias{aeternity}{\ae{ternity}}
\defcitealias{babbage}{Babbage}
\defcitealias{solidityx}{SolidityX}
\defcitealias{bamboo}{Bamboo}
\defcitealias{zen}{Zen}
\defcitealias{obsidian}{Obsidian}
\defcitealias{yul}{\tname{Yul}}
\defcitealias{iele}{\tname{IELE}}
\defcitealias{plutus}{\tname{Plutus}}
\defcitealias{erc20}{ERC20}

\theoremstyle{remark}

\makeatletter 
\def\arcr{\@arraycr}
\makeatother

\definecolor{deepblue}{rgb}{0,0,0.5}
\definecolor{deepred}{rgb}{0.6,0,0}
\definecolor{deepgreen}{rgb}{0,0.5,0}
\definecolor{halfgray}{gray}{0.55}
\definecolor{ipythonframe}{RGB}{207, 207, 207}

\definecolor[named]{ACMBlue}{cmyk}{1,0.1,0,0.1}
\definecolor[named]{ACMYellow}{cmyk}{0,0.16,1,0}
\definecolor[named]{ACMOrange}{cmyk}{0,0.42,1,0.01}
\definecolor[named]{ACMRed}{cmyk}{0,0.90,0.86,0}
\definecolor[named]{ACMLightBlue}{cmyk}{0.49,0.01,0,0}
\definecolor[named]{ACMGreen}{cmyk}{0.20,0,1,0.19}
\definecolor[named]{ACMPurple}{cmyk}{0.55,1,0,0.15}
\definecolor[named]{ACMDarkBlue}{cmyk}{1,0.58,0,0.21}

\definecolor{ckeyword}{HTML}{7F0055}
\definecolor{ccomment}{HTML}{3F7F5F}
\definecolor{cnumber}{HTML}{2A0099}
\definecolor{pblue}{rgb}{0.13,0.13,1}
\definecolor{pgreen}{rgb}{0,0.5,0}
\definecolor{pred}{rgb}{0.9,0,0}
\definecolor{pgrey}{rgb}{0.46,0.45,0.48}

\lstdefinestyle{OCaml}{
  language=Caml,
  identifierstyle=\color{black},
  sensitive=true,
  commentstyle=\color{ccomment}\sffamily,
  string=[b]",
  morekeywords={module, struct, open, include, val, nonrec, sig},  
  showstringspaces=false,
  showspaces=false,
  showtabs=false,
  breaklines=true,
  breakatwhitespace=true,
  lineskip=-0.6pt,
  basewidth={0.54em, 0.4em},%
  basicstyle=\scriptsize\ttfamily,
  keywordstyle={\color{ckeyword}\bfseries},
  ndkeywordstyle={\color{ACMDarkBlue}},
  commentstyle={\color{ccomment}\itshape},
  stringstyle={\color{pgreen}},
  numberstyle={\scriptsize\color{cnumber}\ttfamily},
}

\lstdefinelanguage{Scilla}{ 
  keywords={contract, fun, transition, while, payable, let, in, field,
  send, accept, match, with, type, end, Emp, forall, tfun, event, library},
  ndkeywords={Bool, Address, Map, Pair, ByStr20, Uint128, BNum, Int32,
  Uint32, String},
  identifierstyle=\color{black},
  sensitive=true,
  comment=[l]{//},
  morecomment=[s]{(*}{*)},
  commentstyle=\color{ccomment}\sffamily,
  string=[b]",
  showstringspaces=false,
  showspaces=false,
  showtabs=false,
  breaklines=true,
  morekeywords={contract, returns, return},
  breakatwhitespace=true,
  lineskip=-0.6pt,
  basewidth={0.54em, 0.4em},%
  basicstyle=\small\ttfamily,
  keywordstyle={\color{ckeyword}\bfseries},
  ndkeywordstyle={\color{ACMDarkBlue}},
  commentstyle={\color{ccomment}\itshape},
  stringstyle={\color{pgreen}},
  numberstyle={\scriptsize\color{cnumber}\ttfamily},
  moredelim=[il][\textcolor{pgrey}]{$$},
  moredelim=[is][\textcolor{pgrey}]{\%\%}{\%\%},
  xleftmargin=17pt,
  literate={<<}{{$<$}}1
           {<-}{{$\leftarrow$}}1
           {=>}{{$\Rightarrow$}}1
           {->}{{$\rightarrow$}}1
           {<=}{{$\leq$}}1
}

\definecolor{eclipseStrings}{RGB}{42,0.0,255}
\definecolor{eclipseKeywords}{RGB}{127,0,85}
\colorlet{numb}{magenta!60!black}

\lstdefinelanguage{json}{
    basicstyle=\normalfont\ttfamily,
    commentstyle=\color{eclipseStrings}, 
    stringstyle=\color{eclipseKeywords}, 
    numbersep=8pt,
    showstringspaces=false,
    breaklines=true,
    frame=single,
    lineskip=-0.6pt,
  	basewidth={0.54em, 0.4em},%
 	basicstyle=\scriptsize\ttfamily,
    string=[s]{"}{"},
    comment=[l]{:\ "},
    morecomment=[l]{:"},
    literate=
        *{0}{{{\color{numb}0}}}{1}
         {1}{{{\color{numb}1}}}{1}
         {2}{{{\color{numb}2}}}{1}
         {3}{{{\color{numb}3}}}{1}
         {4}{{{\color{numb}4}}}{1}
         {5}{{{\color{numb}5}}}{1}
         {6}{{{\color{numb}6}}}{1}
         {7}{{{\color{numb}7}}}{1}
         {8}{{{\color{numb}8}}}{1}
         {9}{{{\color{numb}9}}}{1}
}

\definecolor{shadecolor}{gray}{1.00}
\definecolor{ddarkgray}{gray}{0.75}
\definecolor{darkgray}{gray}{0.30}
\definecolor{light-gray}{gray}{0.87}

\newcommand{\ie}{\emph{i.e.}\xspace}
\newcommand{\aka}{\emph{aka}\xspace}

\newcommand{\scilla}{\textsc{Scilla}\xspace}
\newcommand{\zilliqa}{\textsc{Zilliqa}\xspace}

\newcommand{\tname}[1]{\textsc{#1}\xspace}

\newcommand{\ConsC}{\mathsf{Cons}}
\newcommand{\SomeC}{\mathsf{Some}}

\newcommand{\sic}[1]{{\tt #1}}

\begin{document}

\def\sectionautorefname{Sec.}
\def\subsectionautorefname{Sec.}
\def\subsectionautorefname{Sec.}
\def\subsubsectionautorefname{Sec.}
\def\figureautorefname{Fig.}
\def\tableautorefname{Tab.}
\def\equationautorefname{Eq.}

\interfootnotelinepenalty=10000

\newcommand{\mytitle}{Compiling a Higher-Order Smart Contract Language to LLVM}
\title[\mytitle]{\huge{\mytitle}}

\author{Vaivaswatha Nagaraj}
\affiliation{%
  \institution{Zilliqa Research}
}
\email{vaivaswatha@zilliqa.com}

\author{Jacob Johannsen}
\affiliation{%
  \institution{Zilliqa Research}
}
\email{jacob@zilliqa.com}

\author{Anton Trunov}
\affiliation{%
  \institution{Zilliqa Research}
}
\email{anton@zilliqa.com}

\author{George P\^{i}rlea}
\affiliation{%
  \institution{Zilliqa Research}
}
\email{george@zilliqa.com}

\author{Amrit Kumar}
\affiliation{%
  \institution{Zilliqa Research}
}
\email{amrit@zilliqa.com}

\author{Ilya Sergey}
\affiliation{%
  \institution{Yale-NUS College}
}
\affiliation{%
  \institution{National University of Singapore}
}
\email{ilya.sergey@yale-nus.edu.sg}


\begin{abstract}

\scilla is a higher-order polymorphic typed intermediate level language
for implementing smart contracts. In this talk, we describe a \scilla compiler
targeting LLVM, with a focus on mapping \scilla types, values, and its
functional language constructs to LLVM-IR.

The compiled LLVM-IR, when executed with LLVM's JIT framework,
achieves a speedup of about 10x over the reference interpreter on a
typical \scilla contract. This reduced latency is crucial in the
setting of blockchains, where smart contracts are executed as parts of
transactions, to achieve peak transactions processed per second.
Experiments on the Ackermann function achieved a speedup of more than
45x.

This talk is aimed at both programming language researchers looking to
implement an LLVM based compiler for their functional language, as
well as at LLVM practitioners.
\end{abstract}

\maketitle

\section {Introduction}

\scilla ({\bf S}mart {\bf C}ontract {\bf I}ntermediate {\bf L}evel
{\bf La}nguage) \cite{Sergey-al:OOPSLA19} is a functional programming
language in the ML family, designed to enable writing safe smart
contracts. 
\scilla is the main smart contract language of the
\zilliqa blockchain~\cite{zilliqa}. Contracts written in \scilla are
deployed \emph{as-is} (\ie, as source code) and are interpreted during
transaction validation by the nodes participating in the blockchain
consensus (\aka miners).
The project is under active development and is free and
open-source \cite{scilla-gh, scilla-compiler-gh,scilla-vm-gh}.

\begin{figure}[t]
  \vspace{15pt}
  \centering
  \includegraphics[scale=0.53]{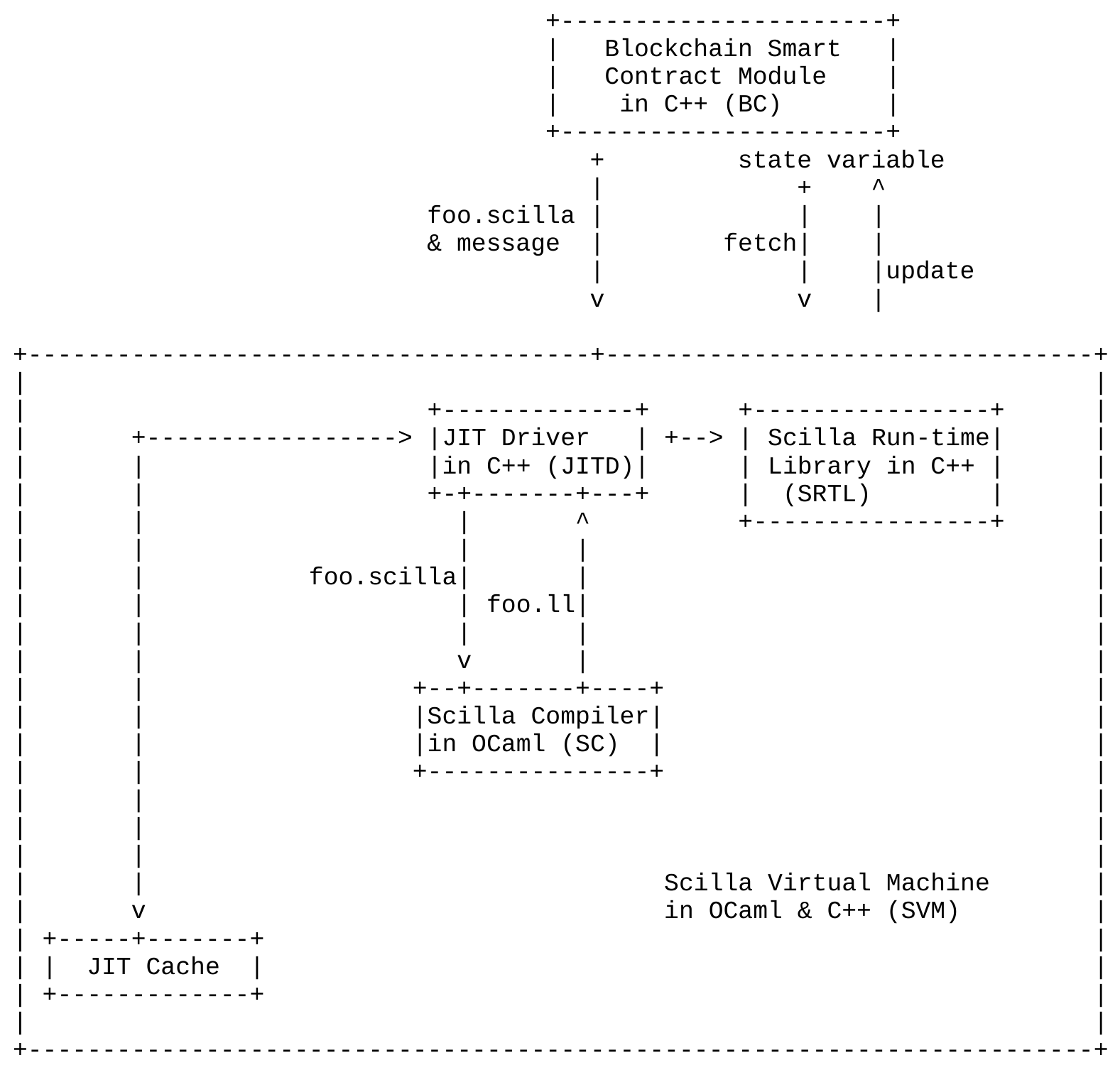}
  \caption{Design of the \scilla Virtual Machine}
  \label{fig:arch}
\end{figure}

The \scilla Virtual Machine (SVM) (\autoref{fig:arch}) aims to eventually
replace the \scilla interpreter with faster execution, thus achieving higher
transaction throughput. This project comprises of the \scilla compiler (SC) which
translates \scilla to LLVM-IR, the \scilla runtime library (SRTL) and the
JIT driver which JIT compiles the LLVM-IR and executes it.
Execution of a \scilla contract may result in accessing the blockchain state,
with such interactions facilitated by the runtime library. The runtime library
also implements \scilla built-in operations.

In the course of this talk, we will briefly describe the phases of our
compiler pipeline (\autoref{sec:pipeline}),
elaborate on mapping the \scilla AST to LLVM-IR (\autoref{sec:llmap}),
present characteristic benchmarks used to evaluate our compiler
(\autoref{sec:eval}), and
share our vision on some desired LLVM features which would facilitate
projects similar to our own (\autoref{sec:nice}).

\section{The Compiler Pipeline}
\label{sec:pipeline}

\scilla's reference interpreter and the accompanying type-checker are
written in OCaml. To take advantage of the developed infrastructure,
we decided to write the compiler also in OCaml, thus enabling re-using
much of the frontend code and the type-checker. The compiler uses
LLVM's OCaml bindings to generate LLVM-IR. 

Serving as a background for the next section, we now briefly discuss
our compiler pipeline. While in this proposal we use textual description,
our talk will make use of examples to illustrate the
transformations performed in each pass.

\subsection {Parsing and Type Checking}

The parser and type checker reuse code from the reference interpreter and the result
is an AST with each identifier annotated with its type. The type annotations are
necessary for code generation later on.

\subsection {Dead Code Elimination}

\scilla follows a whole-program compilation model. This means that along with the
program currently being compiled, we also compile imported libraries. While the
DCE pass is useful in the traditional sense, our primary goal for it here is to
simply eliminate unused imported library functions, to improve compile time.

\subsection {Pattern Match Flattening}

\scilla allows for nested patterns in a match expression. For example,
the expression shown in \autoref{fig:nestedpat} requires checking for a
$\SomeC$ constructor inside an object created with the $\ConsC$
constructor. To efficiently translate pattern match constructs into an
LLVM switch statement, match expressions with nested patterns are
flattened into nested match expressions with flat patterns~\cite{PeytonJones:Book}.

\begin{figure}
\setlength{\abovecaptionskip}{0pt}
\setlength{\belowcaptionskip}{-10pt}
  \begin{center}
  \begin{lstlisting}[language=Scilla,numbers=left,xleftmargin=15pt]
    fun (p : List (Option Int32)) =>
      match p with
      | Nil => z
      | Cons (Some x) xs => x
      | Cons _ _ => z
      end 
  \end{lstlisting}
  \end{center}
  \caption{Nested pattern in a match-expression.}
  \label{fig:nestedpat}
\end{figure}

\subsection {Uncurrying}

As is common in functional programming languages, \scilla functions
and their applications (calls) follow curried
semantics~\cite{WadlerIntro}. This means that a function taking $n$
arguments is actually a function that takes in one argument and
returns a function to process the remaining $n-1$ arguments and
produce the result. The uncurrying pass transforms the AST to have
regular C-like (and LLVM-IR) function call semantics, \ie, functions
may take $n \geq 0$ arguments, and their calls will provide exactly
$n$ arguments. The change in semantics is accompanied by an
optimization pass that analyzes all calls of a function, and when all
of them provide the same $n$ number of arguments, the function is
translated to a function taking $n$ arguments. If it cannot prove so,
then the function is translated to take just one argument, and a
nested function handling the rest. An example: If all calls of the
function \sic{f : T1 -> T2 -> T3} provide two arguments (of type
\sic{T1} and \sic{T2}), then \sic{f} is transformed to have the type
\sic{f : (T1,T2) -> T3}.

At the time of writing this proposal, our uncurrying pass changes the
semantics of the AST to use uncurried semantics, but we do not have the
optimization implemented yet.

\subsection {Monomorphize}
Parametric polymorphism (also called generic programming) is typically
implemented using type-erasure (Java, ML, etc) or monomorphization
(C++, Rust). The former erases all type information in generic code
after type-checking and generates a single copy of the code.
This requires that runtime values be boxed (\ie, values of all types represented
uniformly \cite{MorrisonPoly}) so that they can all be handled uniformly in the
generic code. On the other hand monomorphization specializes the code for each
possible (at runtime) ground type (\ie, types that don't have a type parameter)
\cite{MlToAda}, creating copies of the code. While this results in code
replication, each copy can be efficiently compiled because the runtime values
no longer need to be boxed~\cite{levitypoly}.

A third approach to parametric polymorphism is to wait until
polymorphic code is used, and specialize it then (typically using JIT
compilation). This approach is used in C\#~\cite{GenericsCSharp} and
Ada~\cite{UnboxedPoly}.

\scilla is based on System~F~\cite{Reynolds:SP74,Girard:PhD}--- an expressive
type system with \emph{explicit} type bindings and instantiations. We implement
parametric polymorphism by specializing each polymorphic \scilla expression with
all possible ground types. This uses a higher-order data-flow analysis to track
flow of types into polymorphic expressions~\cite{PoPA}.
As type instantiation may happen in stages for \scilla expressions with several
type parameters, we choose to compile a polymorphic expression as a dynamic dispatch
table, associating ground types with the corresponding monomorphic
version of the expression. This provides for a simpler compilation that saves us
from replicating code outside of the polymorphic expression,
but at the cost of a runtime indirection indexing into the (possibly nested)
dynamic dispatch tables. While this is a mix of existing strategies to implement
parametric polymorphism, we are not aware of one that uses this exact combination.

\subsection {Closure Conversion}

As is typical of higher-order languages, \scilla allows one to pass
functions as arguments to other functions and return them, as well as
to define functions nested inside other functions. This means a
function may have not only its arguments available for use inside, but
also a set of ``free variables'' that are defined outside its
syntactic definition.

The closure conversion pass~\cite{Appel:CPS92} computes the set of free variables
for each function, constructs an aggregate type to hold all the free variables
of the function (conventionally called the function's environment) and lifts
the function definition to the top / global level, as is required for LLVM-IR
generation (where all functions are defined at the top level). The lifted
function takes the environment as an argument.

When functions are used as values in the program, each such variable is now
a pair of (1) pointer to the function definition and (2) the environment.
Function applications (calls) are translated into to calling the function pointer,
with the paired environment being passed as an additional argument.

In our implementation, this pass also flattens the AST by transforming nested
{\bf let-in} expressions into a sequence of assignment statements.

\section {Mapping \scilla to LLVM-IR}
\label{sec:llmap}

The final stage of the compiler pipeline translates the closure converted AST to
LLVM-IR.

\subsection {Type Descriptors}
Although \scilla code is monomorphized by the time we generate LLVM-IR for it
(which means we know the ground type of code and values that we're generating
code for), at times it is necessary for code in the runtime library to operate
on \scilla values. For example, many \scilla contract executions end with sending
out a message (which for the purposes of this discussion is just a composite
\scilla value). The message needs to be serialized into a JSON. It is much simpler
to do the JSON serialization outside in the runtime library rather than generate
code to do that.

Such a scenario gives rise to the requirement that we communicate the type of
a \scilla value to the runtime library. To this end we define type descriptor
structs, defined to be identical in both the runtime library and in the generated
code. Every type in the program being compiled will have a type descriptor built
for it. A global type descriptor table is added into the compiled LLVM module,
and interactions with the runtime library that require the type of a value will
use the type's index in this table.

\subsection { Primitive Types }
\begin{itemize}
  \item {\bf Integer types}: \scilla has signed and unsigned integer types with
  widths 32, 64, 128 and 256. These types are mapped to LLVM integer types, but
  with a wrapper struct type. The wrapper struct enables having a 1-1 relation
  b/w \scilla types and their translated LLVM types (because LLVM does not
  distinguish b/w signed and unsigned integers). For example, \sic{Int32}
  translates to the LLVM type \sic {\%Int32 = type \{ i32 \}}
\item { \bf String types}: \sic{String} and \sic{ByStr} (byte strings
  of arbitrary size) in \scilla translate to LLVM structs containing a
  pointer to their contents and an integer field with the size.
  \item { \bf ByStr{\it X} }: \sic Fixed-sized byte strings (\ie, {\it X}
  is known at compile time) translate to LLVM array type \sic {[{\it X} $\times$ i8]}.
\end{itemize}

\subsection {ADTs}
Algebraic Data Types (also called variants or tagged unions) are composite
types that have one or more ``constructors'' (sum type), each of which allow
defining the value to be a tuple (product type). Figure \ref{fig:adtlist}
shows an integer list defined in \scilla. It has two constructors \sic{Nil}
and \sic{Cons}\footnote{At the time of writing this, \scilla does not support
\emph{user-defined} self-referencing ADTs. \sic{List} is a builtin type.}.

Because different constructors may be used to construct a value of a
specific ADT, ADTs cannot easily be represented unboxed. We represent
ADT values via a pointer to a packed struct containing the actual ADT
value.

For an ADT \sic{tname} with constructors \sic{cname1}, \sic{cname2},
\dots \sic{cnameN}, we define LLVM types 
\sic{\%tname = type \{ i8, \%cname1*, \%cname2*, ... \%cnameN*\}},
\sic{\%cname1 = type <\{ i8, [types in cname1's tuple]\}>}, \sic{\%cname2 = ... } etc.
The LLVM type \sic{\%tname} acts as the placeholder for all values
of this ADT, and a pointer to this type is used to represent the boxed
ADT values. It is always only dereferenced for its first \sic{i8} field,
which is the tag representing the specific constructor used to build
this value. The other fields are defined only for type completeness,
with the idea that, in the future, we can type-check the LLVM-IR at
the level of \scilla types. The tag field is dereferenced at
pattern matches, and based on which branch is taken, the pointer is cast
to a \sic{\%cnameI} pointer and used.

We use packed LLVM structs to represent ADT values so that, when we
build or deconstruct ADT values in the runtime library, we do not have
to bother with architecture specific struct packing / padding.

\subsection {Maps}
Maps are boxed and hence handled using an opaque pointer. We rely on
the runtime library to create a map and perform operations on it.
In the runtime library, maps are defined as \sic{unordered\_map<string, any>}.
This map's value type is \sic{std::any} because it needs to be
able to represent any \scilla type, including nested maps and program specific
user-defined types.

\subsection {Messages}
Message types in \scilla are used to create events, exceptions and
outgoing messages. Message objects are created similar to tuples
in other languages, but with each component being given a name.
In other words, they do not have an explicit predefined type. To enable
the runtime library to work with Message objects, we encode them as
a sequence of triples, each consisting of a name, type descriptor,
and the value itself. The sequence is headed by an integer indicating
the number of fields.

\begin{figure}
  \setlength{\abovecaptionskip}{5pt}
  \setlength{\belowcaptionskip}{-10pt}
  \begin{center}
  \begin{lstlisting}[language=Scilla,numbers=left,xleftmargin=15pt]
    type MyList =
      | Nil
      | Cons of Int32 MyList
  \end{lstlisting}
  \end{center}
  \caption{Defining an integer list in \scilla.}
  \label{fig:adtlist}
\end{figure}

\subsection {Closures}
All \scilla functions are represented as closures, irrespective of whether
they have free variables or not. In the generated LLVM-IR, a closure is
represented by an anonymous struct type \sic{\{ fundef\_sig*, void* \}}
where \sic{fundef\_sig} is the signature of the LLVM function definition.
The \sic{void *} represents the environment pointer. A stronger type
isn't used for the environment pointer because we want to represent
different \scilla functions with the same type, but whose environments
may be different, in a uniform way. By convention, all generated LLVM
functions will take an environment pointer as the first argument. If
the function's return type cannot be ``by value'', then the second
argument will be a stack pointer (``sret'') where the return value
must be stored.

To avoid ABI complexities, generated LLVM functions and the hand written
functions in the runtime library that they may call, all follow the simple
rule that if the value size is larger than two eightbytes~\cite{x86-64-abi},
we define that parameter (or return value) to be passed by reference (stack pointer).

\section {Evaluation}
\label{sec:eval}

The \scilla compiler and the runtime library are still under active
development and are not feature complete. We do however have interesting
\scilla programs and synthetic tests that we can compile and execute end-to-end. 

\begin{enumerate}
  \item \href{https://github.com/Zilliqa/scilla-compiler/blob/master/tests/codegen/contr/simple-map.scilla}{Simple-Map}:
  This is a simple map access contract that is representative of common
  constructs used in a typical \scilla contract. It fetches a map entry,
  does a simple computation on it and stores back the result. While
  the interpreted execution takes about 5ms to execute this, the compiled
  code runs in about 0.5ms.
  \item \href{https://github.com/Zilliqa/scilla-compiler/blob/master/tests/codegen/expr/ackermann_3_7.scilexp}{Ackermann}:
  The Ackermann function, implemented in \scilla takes about 2.2s to compute
  \sic{ackermann(3,7)}. The compiled code computes the same result in about 46ms.
  \item \href{https://github.com/Zilliqa/scilla-compiler/blob/master/tests/codegen/expr/church_nat_stlc.scilexp}{Church Encoding}:
  This code computes Church-encoded numerals with the base type equal
  to \sic{Uint32} and two operations on them: addition and multiplication.
  We compute a term that evaluates to Church-encoded 131099 and then
  we convert it to the native \sic{Uint32}. The interpreted execution
  takes about 0.7s while the compiled execution completes with the
  result in about 4ms.
\end{enumerate}

These experiments do not include the time taken by the \scilla LLVM-IR compiler
or the LLVM JIT compiler, but just the time spent in the actual execution.
(1) We haven't prioritized compile time improvements yet,
so we expect it to be high and not worthy of benchmarking.
(2) In the production environment, we cache compiled code on disk,
and the most frequently used ones in memory. So we expect the impact
of compile times to be low.

\section{LLVM Feature Requests}
\label{sec:nice}

LLVM is a mature framework for building compilers. While it has largely
simplified writing compiler frontends for new languages, in the course of
writing the \scilla compiler we found the following features, which, if
implemented in LLVM, would further simplify and ease projects such as ours.

\subsection{DebugIR}
While attaching source debug information to LLVM-IR is useful for
debugging, at times, for a compiler developer, being able to debug the
LLVM-IR itself can save time. LLVM's \sic{-debug-ir} pass served
exactly this purpose, but has long been removed from the tree.
Considering that a \href{https://reviews.llvm.org/D40778}{previous
  attempt} at reviving this wasn't successful, we decided to adapt the
code into a standalone tool~\cite{debugir}.

\subsection{Shared Definitions with Pre-compiled Code}
Many \scilla operations are implemented in the \scilla runtime library (SRTL).
This requires invoking functions in SRTL from the dynamically (JIT) compiled
\scilla code and passing \scilla values. We currently specify common type and
function definitions both when generating LLVM-IR (in the compiler) as well
as in \sic{.h} files in SRTL. This requires both definitions to be kept in
sync. While this can be automated by compiling the \sic{.h} file to LLVM-IR and
using that during compilation, a more systematic (LLVM provided) method for
doing this may lead to a cleaner approach.

\bibliography{references}

\end{document}